\documentclass[prc,showpacs,preprintnumbers,superscriptaddress,floatfix]{revtex4}
\usepackage{amsfonts}
\usepackage{graphicx}
\usepackage{graphicx,epsfig,latexsym,amssymb}
\usepackage{multirow,amsmath,array,booktabs}
\usepackage{dcolumn}
\usepackage[section]{placeins}
\usepackage{bm}

\begin{document}

\title{Further investigation of the relativistic symmetry by similarity
renormalization group}
\author{Dong-Peng Li, Shou-Wan Chen, Jian-You Guo}
\email[E-mail:]{jianyou@ahu.edu.cn}
\affiliation{School of Physics and Material Science, Anhui University, Hefei 230039,
People's Republic of China}

\begin{abstract}
Following a recent rapid communications[Phys.Rev.C85,021302(R) (2012)], we
present more details on the investigation of the relativistic symmetry by
use of the similarity renormalization group. By comparing the contributions
of the different components in the diagonal Dirac Hamiltonian to the
pseudospin splitting, we have found that two components of the dynamical
term make similar influence on the pseudospin symmetry. The same case also
appears in the spin-orbit interactions. Further, we have checked the
influences of every term on the pseudospin splitting and their correlations
with the potential parameters for all the available pseudospin partners. The
result shows that the spin-orbit interactions always play a role in favor of
the pseudospin symmetry, and whether the pseudospin symmetry is improved or
destroyed by the dynamical term relating the shape of the potential as well
as the quantum numbers of the state. The cause why the pseudospin symmetry
becomes better for the levels closer to the continuum is disclosed.
\end{abstract}

\maketitle

\section{Introduction}

More than 40 years ago, a quasidegeneracy was observed in heavy nuclei
between two single-particle states with the quantum numbers ($n-1,l+2,j=l+3/2
$) and ($n,l,j=l+1/2$). In analogy with the well known spin-symmetry (SS)
breaking for the spin doublets ($n,l,j=l\pm 1/2$), which is one of the most
important concepts for understanding the traditional magic number in atomic
nuclei~\cite{Haxel49,Mayer49}, the pseudospin symmetry (PSS) was proposed by
defining the pseudospin doublets ($\tilde{n}=n-1$, $\tilde{l}=l+1$, $j=%
\tilde{l}\pm 1/2$)~\cite{Hecht69,Arima69}. The introduction of the PSS
concept has explained numerous phenomena in nuclear structure including
deformation \cite{Bohr82}, superdeformation \cite{Dudek87}, identical bands
\cite{Nazar90}, and magnetic moment \cite{Trolt94}. Especially for the
magic number change in exotic nuclei, the spin and pseudospin symmetries
are pointed out to play important roles. For instance, the $N=28$ shell
closure disappears due to the quenching of the spin-orbit splitting for the $%
{\nu }1f$ spin doublets \cite{Gaudefroy06,Bastin07,Tarpanov08,Moreno10}, and
the $Z=64$ sub-shell closure is closely related to the restoration of PSS
for the ${\pi }2\tilde{p}$ and ${\pi }1\tilde{f}$ pseudospin doublets \cite%
{Nagai81,Long07,Long09}. Because of these successes, there have been
comprehensive efforts to understand the origin of this symmetry as well as
its breaking mechanism. Based on the single-particle Hamiltonian of the
oscillator shell model, Bahri et al. indicated that the origin of PSS is
connected with a special ratio in the strength of the spin-orbit and
orbit-orbit interactions, and the ratio can be partly explained by the
relativistic mean field theory \cite{Bahri92}. Blokhin et al. introduced a
helicity unitary transformation which can map a normal state ($l,s$) to a pseudo
state ($\tilde{l},\tilde{s}$) \cite{Blokh95}. A substantial
progress was achieved in Ref.~\cite{Ginoc97}, where the relativistic feature
of PSS was recognized. The pseudo-orbital angular momentum $\tilde{l}$ is
nothing but the orbital angular momentum of the lower component of the Dirac
spinor, and the equality in magnitude but difference in sign of the scalar
potential $S$ and vector potential $V$ was suggested as the exact PSS limit.
In a more general condition, the exact PSS is satisfied in the Dirac
equation when the sum of the scalar $S$ and vector $V$ potentials is equal
to a constant \cite{Meng98}. Moreover, the PSS in real nuclei was shown in
connection with the competition between the pseudocentrifugal barrier and
the pseudospin-orbital potential \cite{Meng99,Sugawara00}. Even with these
excellent work mentioned above, there is still much attention on the cause
of splitting for the reason that the exact PSS cannot be met in real nuclei.
In Refs. \cite{Alber01,Alber02,Lisboa10}, it was pointed out that the
observed pseudospin splitting arises from a cancelation of the several
energy components, and the PSS in nuclei has a dynamical character. A
similar conclusion was reached in Refs. \cite{Marco01,Marco08}. In addition,
it was noted that, unlike the spin symmetry, the pseudospin breaking cannot
be treated as a perturbation of the pseudospin-symmetric Hamiltonian\cite%
{Liang11}. The nonperturbation nature of PSS has also been mentioned in Ref.
\cite{Gonoc11}. Further, the supersymmetric description of PSS was presented
for the spherical nuclei and axially deformed nuclei \cite%
{Leviatan04,Typel08,Leviatan09}. In a very recent paper~\cite{Liang13},
supersymmetric quantum mechanics and similarity renormalization group (SRG)
are used as the critical tools for understanding the origin of PSS and its
breaking mechanism, and the cause why the PSS becomes better for the levels
closer to the continuum is discussed in a quantitative way at the
nonrelativistic limit. This symmetry is also checked in the resonant states
~\cite{Guo05,Guo06} with similar features to bound states indicated in Ref.~\cite{Lu12}.

Regardless of these pioneering studies, the origin of PSS has not been fully
understood in the relativistic framework. Recently, we have examined the PSS
by use of the similarity renormalization group and shown explicitly the
relativistic origin of this symmetry \cite{Guo121}. We have also studied the
relativistic effect of this symmetry \cite{Guo122}. However, we have not
researched the dependence of PSS on the shape of the potential and the
quantum numbers of the states by using this Hamiltonian. Although the
correlation of the energy splitting of pseudospin partners with the nuclear
potential parameters has been investigated by solving the Dirac equation
with Woods-Saxon scalar and vector radial potentials \cite{Alber01,Alber02},
we do not know the relationship between the splitting of every component and
the shape of the potential, which is particularly important to reveal the
origin of PSS. In this paper, we explore the dependence of the PSS on the
shape of the potential and the quantum numbers of the states by using this
Hamiltonian obtained in Ref. \cite{Guo121} in order to disclose the
influence of every component, especially those relating the dynamical effect
and the spin-orbit interactions.

\section{Formalism}

For simplicity, we sketch our formalism with the following Dirac
Hamiltonian:
\begin{equation}
H_{D}=\left(
\begin{array}{cc}
M+\Sigma (r) & -\frac{d}{dr}+\frac{\kappa }{r} \\
\frac{d}{dr}+\frac{\kappa }{r} & -M+\Delta (r)%
\end{array}%
\right) ,  \label{hamilton}
\end{equation}%
where $\Sigma (r)=V(r)+S(r)$ and $\Delta (r)=V(r)-S(r)$ denote the
combinations of the scalar potential $S(r)$ and the vector potential $V(r)$
and $\kappa $ is defined as $\kappa =(l-j)(2j+1)$. To extract the different
components from $H_{D}$, the SRG is used to transform the Dirac Hamiltonian
into a block-diagonal form. The details can be referred to the literature
\cite{Guo121}. The diagonal Dirac Hamiltonian is written as%
\begin{equation}
H_{D}=\left(
\begin{array}{cc}
H_{P}+M & 0 \\
0 & -H_{P}^{C}-M%
\end{array}%
\right) ,  \label{diagDirac}
\end{equation}%
where
\begin{eqnarray}
H_{P} &=&\Sigma \left( r\right) +\frac{p^{2}}{2M}-\frac{1}{2M^{2}}\left(
Sp^{2}-S^{\prime }\frac{d}{dr}\right) -\frac{\kappa }{r}\frac{\Delta
^{\prime }}{4M^{2}}+\frac{\Sigma ^{\prime \prime }}{8M^{2}}  \notag \\
&&+\frac{S}{2M^{3}}\left( Sp^{2}-2S^{\prime }\frac{d}{dr}\right) +\frac{%
\kappa }{r}\frac{S\Delta ^{\prime }}{2M^{3}}-\frac{{\Sigma ^{\prime }}%
^{2}-2\Sigma ^{\prime }\Delta ^{\prime }+4S\Sigma ^{\prime \prime }}{16M^{3}}%
-\frac{p^{4}}{8M^{3}}+O\left( \frac{1}{M^{4}}\right)  \label{Diracp}
\end{eqnarray}%
is an operator describing Dirac particle with $p^{2}=-\frac{d^{2}}{dr^{2}}+%
\frac{\kappa \left( \kappa +1\right) }{r^{2}}$, and $H_{P}^{C}$ is the
charge-conjugation of $H_{P}$ \cite{Lisboa10,Zhou03}. The charge-conjugation
operator is given by $C=i\gamma ^{2}K$, where $K$ is the complex conjugation
operator~\cite{Itzykson80}. The primes and the double primes in $H_{P}$
respectively denote first- and second-order derivatives with respect to $r$.
From Eq.(\ref{Diracp}), we see that the present transformation avoids all
the drawbacks in the usual decoupling~\cite{Alber01,Alber02}, and $H_{P}$
holds the form of Schr\"{o}dinger-like operator.

\section{The numerical calculations and discussions}

With the formalism represented in the previous section, we explore the
origin of PSS by use of $H_{P}$. To convince the reliability of the present
calculations, we first check the convergence of the expansion in Eq.~(\ref%
{Diracp}). For convenience in the numerical computations, a Woods-Saxon type
potential is adopted for $\Sigma (r)$ and $\Delta (r)$, i.e., $\Sigma
(r)=\Sigma _{0}f(a_{\Sigma },r_{\Sigma },r)$ and $\Delta (r)=\Delta
_{0}f(a_{\Delta },r_{\Delta },r)$ with
\begin{equation}
f(a,R,r)=\frac{1}{1+\exp \left( \frac{r-R}{a}\right) }\text{.}
\end{equation}%
which is realistic enough to be applied to nuclei although it is not a full
self-consistent relativistic potential derived from meson fields \cite%
{Alber02}. Using this potential for $\Sigma (r)$ and $\Delta (r)$,
there are six parameters, i.e., the central depths, $\Sigma _{0}$ and $%
\Delta _{0}$, two radii, and two diffuseness parameters. Following Refs.
\cite{Alber02}, the same radius $R$ and surface diffuseness $a$ are
set for the both potentials $\Sigma (r)$ and $\Delta (r)$, and $\Sigma _{0}$%
, $\Delta _{0}$, $a$, and $R$ are determined by fitting the neutron spectra
of $^{208}$Pb with the fitted results: $\Sigma _{0}=-66$ MeV, $\Delta
_{0}=650 $ MeV, $a=0.6$ fm, and $R=7$ fm. With the values for the
parameters, the energy spectra of $H_{P}$ are obtained and listed in Table I
in comparisons with the solutions of the original Dirac equation. From Table
I, it can be seen that the deviations between the nonrelativistic limit (the
second column) and the exact relativistic case (the fifth column) are very
large, i.e., the relativistic effect is apparent in the present system. With
the increasing perturbation order, the calculated result is closer to the
exact relativistic one. When $H_{P}$ is approximated to the order $1/M^{3}$,
the calculated spectra are considerably agreeable with those from the exact
relativistic calculations. These indicate that the convergence of the
present expansion is satisfactory, and it is reasonable enough to probe
the PSS hidden in Dirac Hamiltonian in terms of the operator $H_{P}$.
\begin{table}[tbp]
\caption{The energy spectra of $H_{p}$ for all the available bound states.
The second column indicates that $H_{p}$ is approximated to the
nonrelativistic limit. The third and fourth columns indicate that $H_{p}$ is
approximated to the orders $1/M^{2}$ and $1/M^{3}$, respectively. For comparisons, the exact relativistic spectra are displayed in the fifth column.}%
\begin{tabular}{c|c|c|c|c}
\hline\hline
$i$ & Non & $1/M^2$(MeV) & $1/M^3$(MeV) & Exa(MeV) \\ \hline
$1s_{1/2}$ & -61.123 & -59.826 & -59.408 & -59.226 \\
$2s_{1/2}$ & -48.353 & -43.689 & -42.221 & -41.609 \\
$3s_{1/2}$ & -30.377 & -21.630 & -19.118 & -18.371 \\
$1p_{3/2}$ & -56.332 & -53.843 & -53.088 & -52.782 \\
$2p_{3/2}$ & -40.463 & -34.018 & -32.104 & -31.416 \\
$3p_{3/2}$ & -20.738 & -10.783 & -8.259 & -7.706 \\
$1p_{1/2}$ & -56.332 & -53.636 & -52.723 & -52.282 \\
$2p_{1/2}$ & -40.463 & -33.608 & -31.446 & -30.625 \\
$3p_{1/2}$ & -20.738 & -10.289 & -7.599 & -7.012 \\
$1d_{5/2}$ & -50.534 & -46.710 & -45.631 & -45.252 \\
$2d_{5/2}$ & -31.981 & -23.908 & -21.675 & -21.012 \\
$1d_{3/2}$ & -50.534 & -46.194 & -44.738 & -44.071 \\
$2d_{3/2}$ & -31.981 & -23.088 & -20.423 & -19.585 \\
$1f_{7/2}$ & -43.842 & -38.626 & -37.282 & -36.898 \\
$2f_{7/2}$ & -23.031 & -13.667 & -11.312 & -10.771 \\
$1f_{5/2}$ & -43.842 & -37.651 & -35.626 & -34.790 \\
$2f_{5/2}$ & -23.031 & -12.389 & -9.483 & -8.788 \\
$1g_{9/2}$ & -36.346 & -29.767 & -28.256 & -27.936 \\
$2g_{9/2}$ & -13.770 & -3.770 & -1.636 & -1.304 \\
$1g_{7/2}$ & -36.346 & -28.183 & -25.621 & -24.714 \\
$1h_{11/2}$ & -28.128 & -20.307 & -18.757 & -18.559 \\
$1h_{9/2}$ & -28.128 & -17.976 & -14.982 & -14.128 \\
$1i_{13/2}$ & -19.263 & -10.429 & -8.995 & -8.955 \\
$1i_{11/2}$ & -19.263 & -7.259 & -4.040 & -3.371 \\ \hline\hline
\end{tabular}%
\end{table}

For analyzing the PSS, we decompose $H_{P}$ into the eight components: $%
\Sigma \left( r\right) +\frac{p^{2}}{2M}$, $-\frac{1}{2M^{2}}\left(
Sp^{2}-S^{\prime }\frac{d}{dr}\right) $, $\frac{S}{2M^{3}}\left(
Sp^{2}-2S^{\prime }\frac{d}{dr}\right) $, $-\frac{\kappa }{r}\frac{\Delta
^{\prime }}{4M^{2}}$, $\frac{\kappa }{r}\frac{S\Delta ^{\prime }}{2M^{3}}$, $%
\frac{\Sigma ^{\prime \prime }}{8M^{2}}$, $-\frac{{\Sigma ^{\prime }}%
^{2}-2\Sigma ^{\prime }\Delta ^{\prime }+4S\Sigma ^{\prime \prime }}{16M^{3}}
$, $-\frac{p^{4}}{8M^{3}}$, which are respectively labeled as $%
O_{1},O_{2},\ldots ,O_{8}$. $O_{1}$ corresponds to the operator describing
Dirac particle in the nonrelativistic limit. $O_{2}$ and $O_{3}$ are related
to the dynamical effect. The spin-orbit interactions are reflected in the $%
O_{4}$ and $O_{5}$. In this decomposition $O_{i}(i=1,2,\ldots ,8)$ is
Hermitian. Hence, it is easy to calculate the contribution of each component
to the pseudospin splitting, which is helpful to disclose the origin of
PSS. The contribution of $O_{i}$ to the level $E_{k}$ is calculated by the
formula $\left\langle k\right\vert O_{i}\left\vert k\right\rangle =\int \psi
_{k}^{\ast }O_{i}\psi _{k}d^{3}\vec{r}$, where $\psi _{k}$ is a eigenvector
of the $k$-state.

To clarify whether the quality of PSS originates mainly from the competition
of the dynamical effect and the spin-orbit interactions, we compare the
contributions of the dynamical components ($O_{2}$ and $O_{3}$) and the
spin-orbit interactions ($O_{4}$ and $O_{5}$) to the energy splitting of
pseudospin partner, and their correlations with the shape of potential and
the quantum numbers of the state.

In Fig.1, we show the variation of the pseudospin energy splitting with the
surface diffuseness $a$ for the doublets $(3s_{1/2},2d_{3/2})$ and $%
(3p_{3/2},2f_{5/2})$, where the "dynam1" and "dynam2" present respectively
the contributions of $O_{2}$ and $O_{3}$ to the energy splitting, and the
"spin-orb1" and "spin-orb2" present respectively the contributions of $O_{4}$
and $O_{5}$ to the energy splitting. For guiding eyes, the total energy
splitting is plotted as "total". The same labels are adopted in the
following figures 2 and 3. From Fig.1, we can see that the pseudospin
splitting caused by the component $O_{i}$ $(i=2,3,4,5)$ is insensitive to $a$%
. The sensitivity of total energy splitting to $a$ origins mainly from the
contribution of the nonrelativistic part, which will be seen in the latter
discussions. Over the range of $a$ under consideration, these components
relating the spin-orbit interactions always play a role in favor of the PSS,
while the dynamical effect depends on the particular pseudospin doublet we
are considering. For the doublet $(3s_{1/2},2d_{3/2})$, the pseudospin
splitting is added by the contributions of the dynamical components, while
for the doublet $(3p_{3/2},2f_{5/2})$, the pseudospin splitting is reduced
by the contributions of the dynamical components. The same case also appears
in the energy splitting varying with the depth of potential well, which is
plotted in Fig.2. Over the range of $\Sigma _{0}$ under consideration, the
contributions of $O_{4}$ and $O_{5}$ to the energy splitting are negative,
i.e., improve the PSS, while the contributions of $O_{2}$ and $O_{3}$ to the
energy splitting are positive for the $(3s_{1/2},2d_{3/2})$, and negative
for the $(3p_{3/2},2f_{5/2})$. The pseudospin energy splitting varying with
the radius $R$ is depicted in Fig.3. Similar to the previous two figures, the PSS is improved by the spin-orbit interactions for all the pseudospin partners considered here. But for the dynamical components, which contribute the pseudospin splitting evolving from a negative value to a positive value and inverting the sign of the energy splitting with the increasing of $R$. The trend of total energy
splitting with $R$ is consistent with that caused by the dynamical
components. Especially for the $(3p_{3/2},2f_{5/2})$, the dynamical effect
is sensitive to $R$. Whether the PSS is improved or destroyed by the
dynamical components depends on the shape of potential.

From Figs.1-3, we have noticed that the contribution of $O_{2}$ to the
pseudospin splitting is similar to that of $O_{3}$ except for the extent of
splitting. When a splitting value caused by $O_{2}$ is negative, that by $%
O_{3}$ is also negative. With the change of the potential parameters, the
variation of the splitting caused by $O_{2}$ is consistent with that by $%
O_{3}$, even the position of the splitting appearing inversion is same. The
similar case also appears in the spin-orbit interactions ($O_{4}$ and $O_{5}$%
). Accordingly, to compare the contributions of these terms with different
physical effect to the pseudospin splitting, we combine $O_{2}$ and $O_{3}$
as a dynamic term, and $O_{4}$ and $O_{5}$ as a spin-orbit coupling term. As
the influences of $O_{6}$, $O_{7}$, and $O_{8}$ on the PSS are weak, we
combine them as an other term. With these combinations, we compare the
dependence of the contribution of every term to the pseudospin splitting on
the shape of potential and the quantum numbers of the state for all the
available pseudospin partners.

In Figs.4-6, we display the variation of the pseudospin energy splitting
with the surface diffuseness $a$ for the $(2p_{3/2},1f_{5/2})$, $%
(2d_{5/2},1g_{7/2})$, $(3s_{1/2},2d_{3/2})$, $(3p_{3/2},2f_{5/2})$, $%
(2f_{7/2},1h_{9/2})$, and $(2g_{9/2},1i_{11/2})$. The relativistic and
nonrelativistic pseudospin splittings are sensitive to $a$. The trend of
total energy splitting with $a$ is similar to the nonrelativistic case. The
contributions from the other parts in $H_{P}$ almost do not alter with $a$.
The variation of total energy splitting with $a$ origins mainly from the
nonrelativistic part. Nevertheless, the relativistic PSS is significantly
improved, which comes mainly from the spin-orbit interactions and the
dynamical effect. The improvement from the spin-orbit interactions increases
with the increasing orbital angular momentum for the states with the same
radial quantum number. Different from the spin-orbit interactions, the
dynamical term destroys PSS for the deeply bound states $(2p_{3/2},1f_{5/2})$%
, $(2d_{5/2},1g_{7/2})$, $(3s_{1/2},2d_{3/2})$, and $(2f_{7/2},1h_{9/2})$,
while improves PSS for the loosely bound states $%
(3p_{3/2},2f_{5/2})$ and $(2g_{9/2},1i_{11/2})$. The contributions of $O_{6}$%
, $O_{7}$, and $O_{8}$ to the pseudospin splitting are negligible.

The pseudospin energy splitting varying with the depth of potential is
exhibited in Figs.7-9 for these partners shown in Figs.4-6. The potential
depth $\Sigma _{0}$ almost does not affect the splitting contributed by the
nonrelativistic part regardless of the splitting is serious. The evolution
of pseudospin energy splitting with $\Sigma _{0}$ is mostly dominated by the
spin-orbit interactions and the dynamical effect. Over the range of $\Sigma
_{0}$ here, the improvement from the spin-orbit interactions increases with
the increasing orbital angular momentum for the states with the same radial
quantum number, which is similar to the case with $a$. Whether the PSS is
improved or destroyed by the dynamical term relating the particular
pseudospin doublet. For the pseudospin partners $(2p_{3/2},1f_{5/2})$, $%
(2d_{5/2},1g_{7/2})$, $(3s_{1/2},2d_{3/2})$, and $(2f_{7/2},1h_{9/2})$, the
contribution of the dynamic term enlarges the pseudospin energy splitting.
However for the $(3p_{3/2},2f_{5/2})$ and $(2g_{9/2},1i_{11/2})$, the
contribution of the dynamical term reduces the pseudospin energy splitting,
and the dynamical effect becomes an improvement to the PSS. The improvement
increases as the potential depth becomes shallow. Compared with the
spin-orbit interactions, the dynamic effect is more sensitive to $\Sigma
_{0} $. Especially for the loosely bound states $(3p_{3/2},2f_{5/2})$ and $%
(2g_{9/2},1i_{11/2})$, the PSS improved with the gradually shallow potential
well originates mainly from the dynamic effect.

Besides the $a$ and $\Sigma_{0}$, the relationship of the pseudospin
splitting and the radius $R$ is more interesting. In Figs.10-12, we show the
pseudospin energy splitting varying with $R$ for all the available
pseudospin partners. The nonrelativistic energy splitting decreases with the
increasing of $R$, which is opposite with total energy splitting with a
little exception (e.g., $(2p_{3/2},1f_{5/2})$ ). The increasing of total
energy splitting with $R$ arises from the contributions of the dynamical
term and the spin-orbit interactions. Similar to the preceding case, the
spin-orbit interactions always improve the PSS, but the improvement becomes
weaker with the increasing of $R$. For the pseudospin partners $%
(2p_{3/2},1f_{5/2})$ and $(2d_{5/2},1g_{7/2})$, the dynamical effect is insensitive to $R$, and the variation of total energy splitting with $%
R$ comes mainly from the the spin-orbit interactions. However for the other
parters, the dynamic effect improves the PSS when $R$ is small. With the
increasing of $R$, the pseudospin splitting coming from the dynamic effect
appears inversion, from an improvement to a breaking. Together, they create
a variation of total energy splitting with $R$. But the sensitivity to $R$
is different. For example the $(3p_{3/2},2f_{5/2})$ partner, the variation
of total energy splitting with $R $ origins mainly from the dynamic effect,
the splitting from the spin-orbit interactions is almost independent of $R$.

Throughout Figs.4-12, the spin-orbit interactions and the dynamical effect
play the key roles in influencing the PSS. Their contributions to the
pseudospin energy splitting are correlated with the shape of the potential
and the quantum numbers of the state. Compared with the spin-orbit
interactions, the dependence of the dynamical effect on the shape of
potential is more sensitive. Over the range of the potential parameters
considered here, the spin-orbit interactions always play a role in improving
the PSS. The improvement of the spin-orbit interactions to the PSS increases
for these levels closer to the continuum. However for the dynamical effect,
it relates the shape of the potential as well as the quantum numbers of the
state. For the deeply bound levels, the contribution of the dynamical term
is a breaking of the PSS, while for the levels close to the continuum, the
contribution of dynamical term becomes an improvement to the PSS. These have
explained the reason why the levels are closer to the continuum, the PSS is
better. In short, the quality of PSS is related to the shape of the potential and the quantum numbers of the state as well as the relativistic effect.

\section{Summary}

In summary, by using the Hamiltonian obtained from the usual Dirac
Hamiltonian by the similarity renormalization group, we have researched in
details the origin of the relativistic symmetry in nuclei. By comparing the
contribution of the different components in the Dirac Hamiltonian $H_P$ to
the pseudospin energy splitting, and their relation to the shape of
potential, it is found that two dynamic components make similar effect on
the PSS. The same case also appears in the spin-orbit interactions. Further,
we have checked the contribution of every term in the Hamiltonian $H_P$ to
the pseudospin splitting, and their correlations with the potential
parameters for all the available pseudospin partners. The results show that
the spin-orbit interactions and the dynamical effect play the major role in
influencing the PSS. Their contributions to the pseudospin energy splitting
are correlated with the shape of the potential and the quantum numbers of
the state. Compared with the spin-orbit interactions, the dependence of the
dynamical effect on the shape of potential is more sensitive. Over the range
of the potential parameters considered here, the spin-orbit interactions
always improves the PSS. For these levels closer to the continuum, the
improvement of the spin-orbit interactions to the PSS is more obvious.
However for the dynamical effect, whether the PSS is improved or destroyed
by the dynamical term relating the shape of potential and the quantum
numbers of the state. For the deeply bound levels, the contribution of the
dynamical term to the pseudospin splitting is against that of the spin-obit
interactions, the quality of PSS originates mainly from the competition of
the dynamical effects and the spin-orbit interactions. However for the
levels close to the continuum, the contribution of the dynamical term
reduces the pseudospin splitting just like the spin-orbit interactions.
These have explained the reason why the levels are closer to the continuum,
the PSS is better. The quality of PSS is related to the shape of the
potential and the quantum numbers of the state as well as the relativistic
effect.

%\begin{acknowledgments}
This work was partly supported by the National Natural Science Foundation of
China under Grants No. 10675001, No. 11175001, and No. 11205004; the Program
for New Century Excellent Talents in University of China under Grant No.
NCET-05-0558; the Excellent Talents Cultivation Foundation of Anhui Province
under Grant No. 2007Z018; the Natural Science Foundation of Anhui Province
under Grant No. 11040606M07; the Education Committee Foundation of Anhui
Province under Grant No. KJ2009A129; and the 211 Project of Anhui
University. %\end{acknowledgments}

\newpage

\begin{figure}[tbp]
\includegraphics[width=8cm]{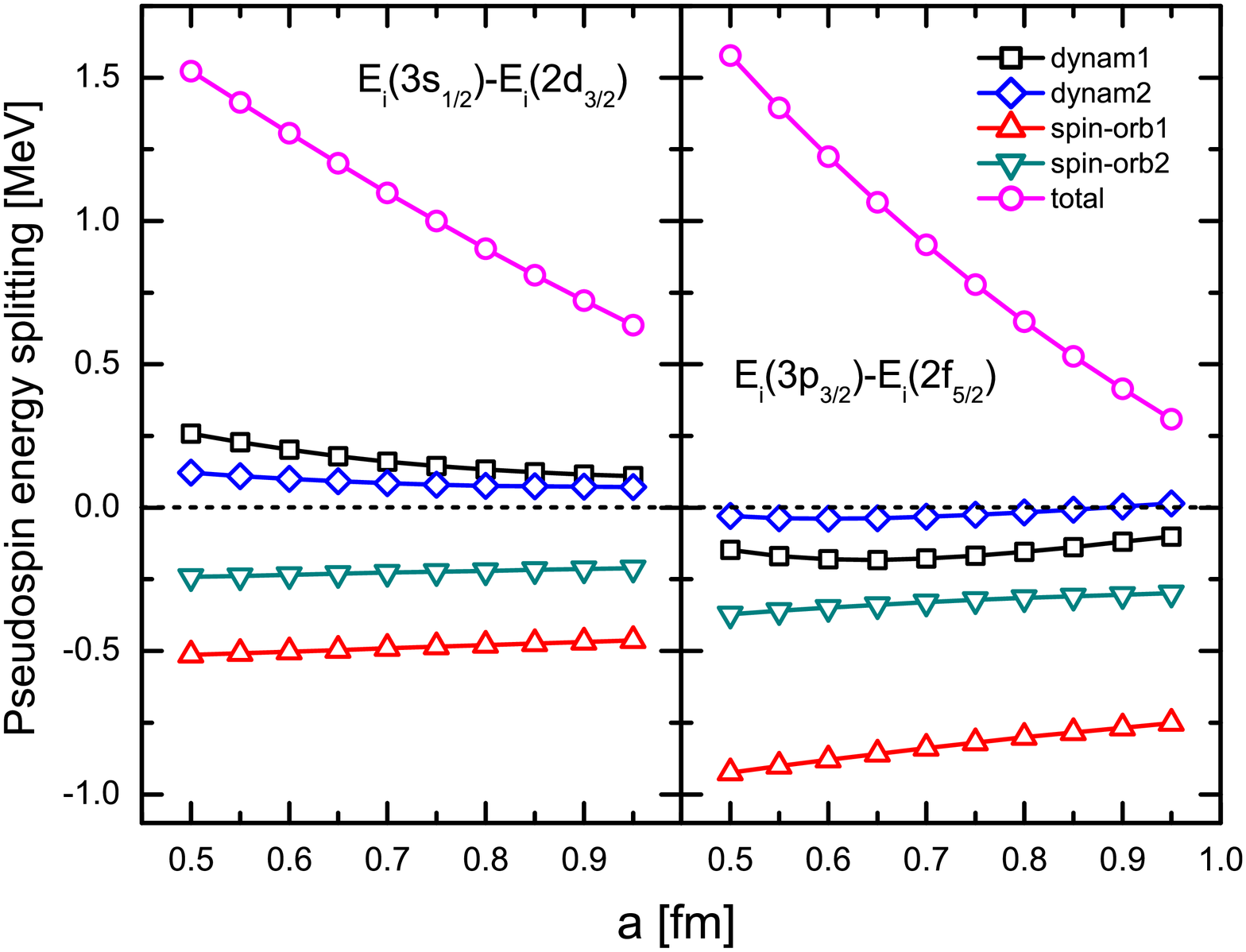}
\caption{(Color online) Comparisons of the contributions of the components
relating the dynamical term and the spin-orbit interactions to the
pseudospin energy splitting and their correlations with the surface
diffuseness $a$ for the $(3s_{1/2},2d_{3/2})$ and $(3p_{3/2},2f_{5/2})$
partners, where the "dynam1(spin-orb1)" and "dynam2 (spin-orb2)" correspond
respectively to the $1/M^2$ order and the $1/M^3$ order perturbations for
the dynamical components (the spin-orbit interactions). For guiding eyes,
the total pseudospin energy splitting is plotted as "total".}
\end{figure}

\begin{figure}[tbp]
\includegraphics[width=8cm]{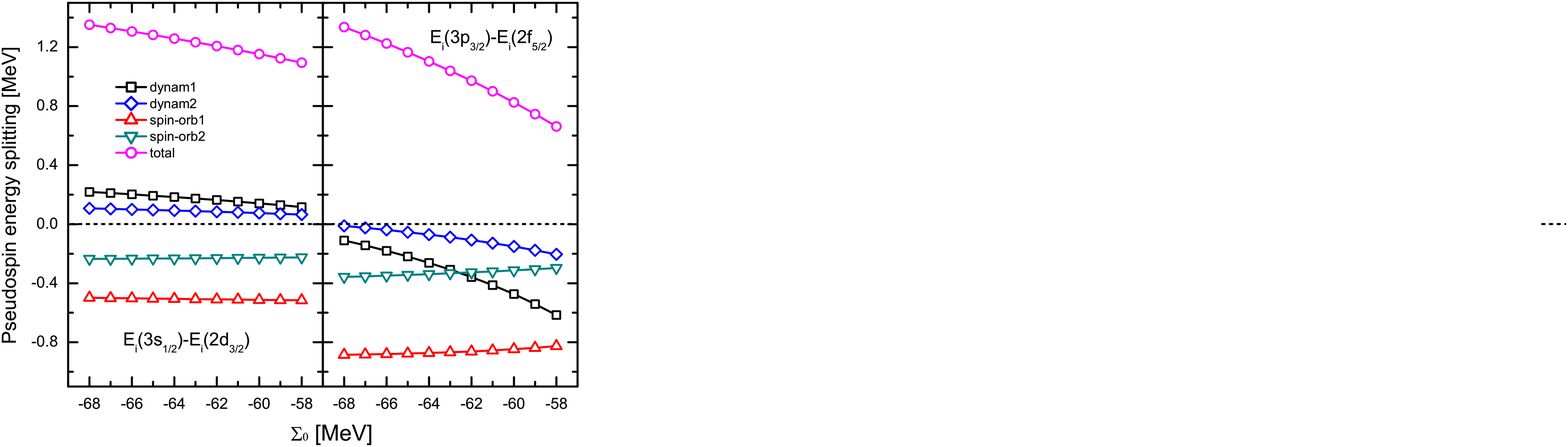}
\caption{(Color online) The same as Fig.1, but with the depth $\Sigma_0$ of
the Woods-Saxon potential.}
\end{figure}

\begin{figure}[tbp]
\includegraphics[width=8cm]{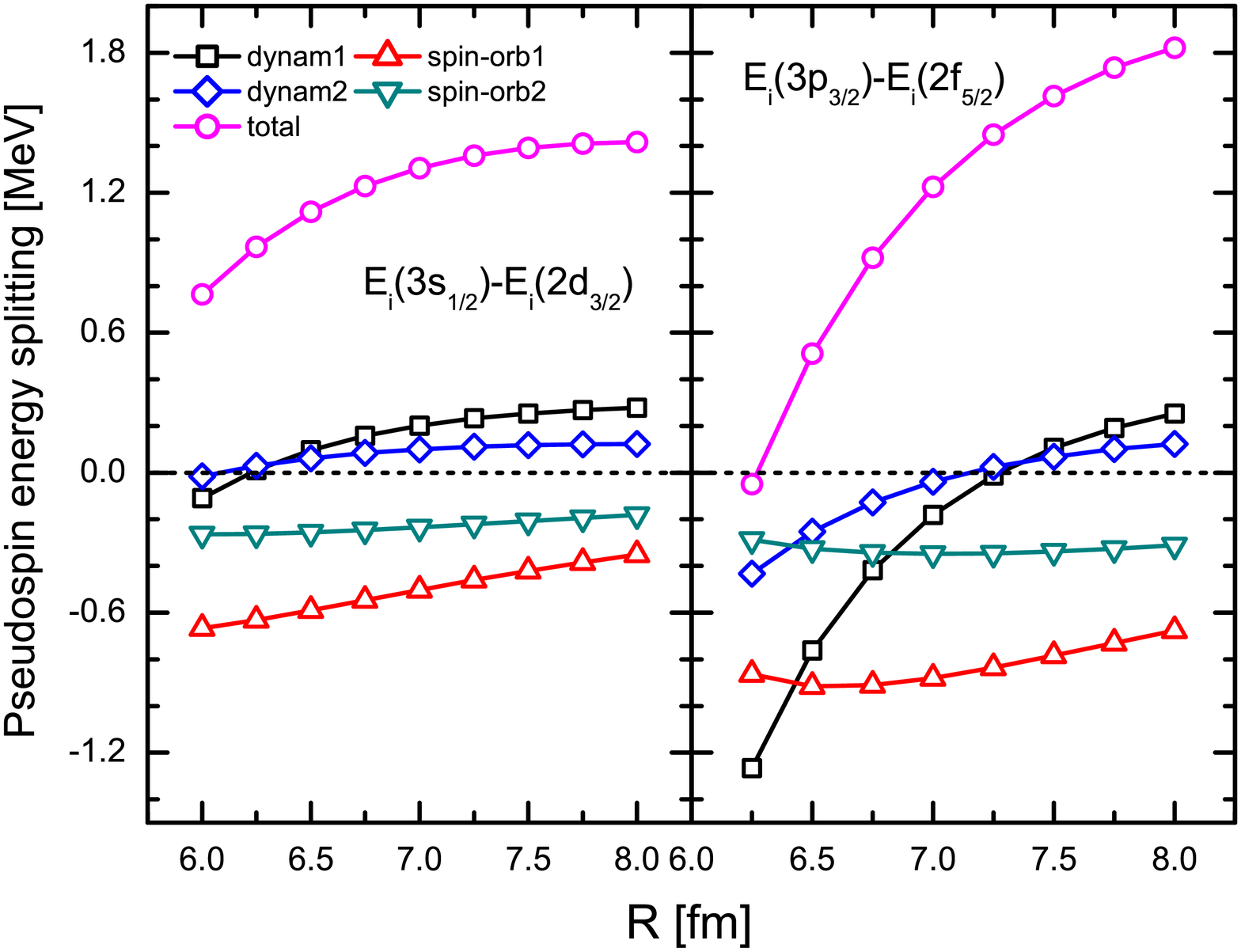}
\caption{(Color online) The same as Fig.1, but with the radius $R$ of the
Woods-Saxon potential.}
\end{figure}

\begin{figure}[tbp]
\includegraphics[width=8cm]{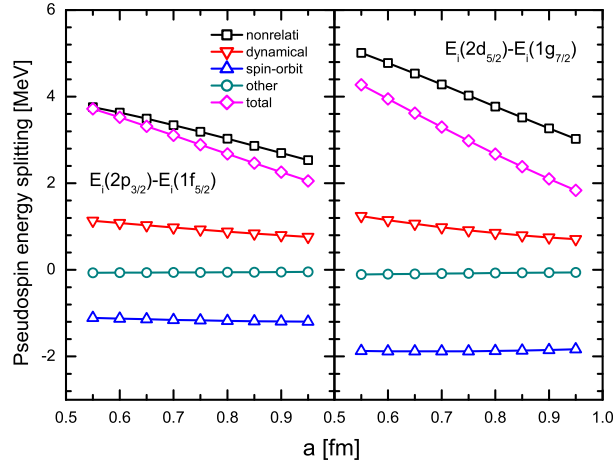}
\caption{(Color online) Comparisons of the contributions of all the terms in
$H_{P}$ to the pseudospin energy splitting and their correlations with the
surface diffuseness $a$ for the $(2p_{3/2},1f_{5/2})$ and $%
(2d_{5/2},1g_{7/2})$ partners, where "nonrelati" denotes the result in the
nonrelativistic limit, "dynamical (spin-orbit)" denotes the data contributed
by the dynamical term (the spin-orbit interactions), "other" marks a
combination of the contributions of $O_{6}$, $O_{7}$, and $O_{8}$ to the
pseudospin splitting, and "total" labels the total pseudospin energy
splitting. }
\end{figure}

\begin{figure}[tbp]
\includegraphics[width=8cm]{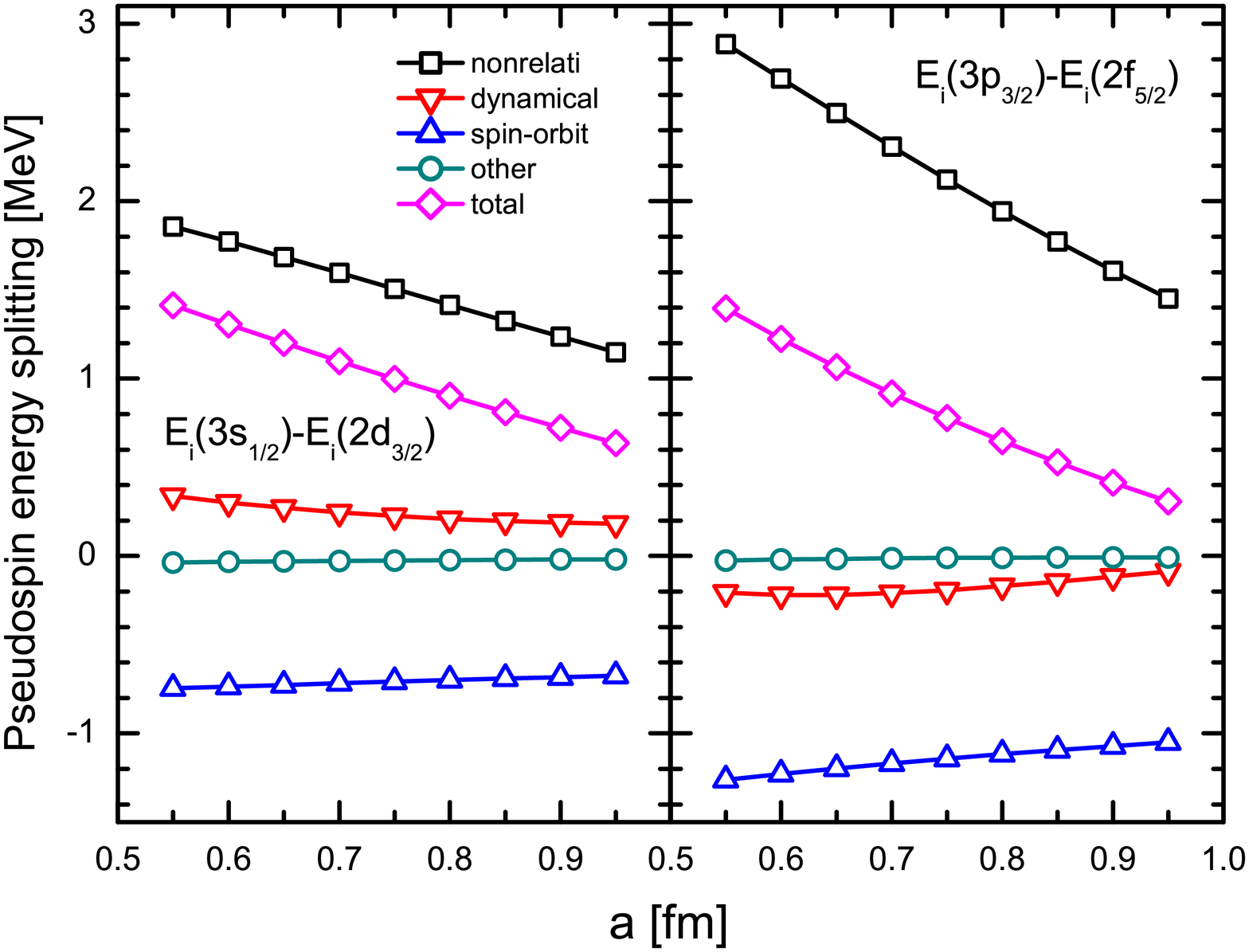}
\caption{(Color online) The same as Fig.4, but for the $(3s_{1/2},2d_{3/2})$
and $(3p_{3/2},2f_{5/2})$ partners. }
\end{figure}

\begin{figure}[tbp]
\includegraphics[width=8cm]{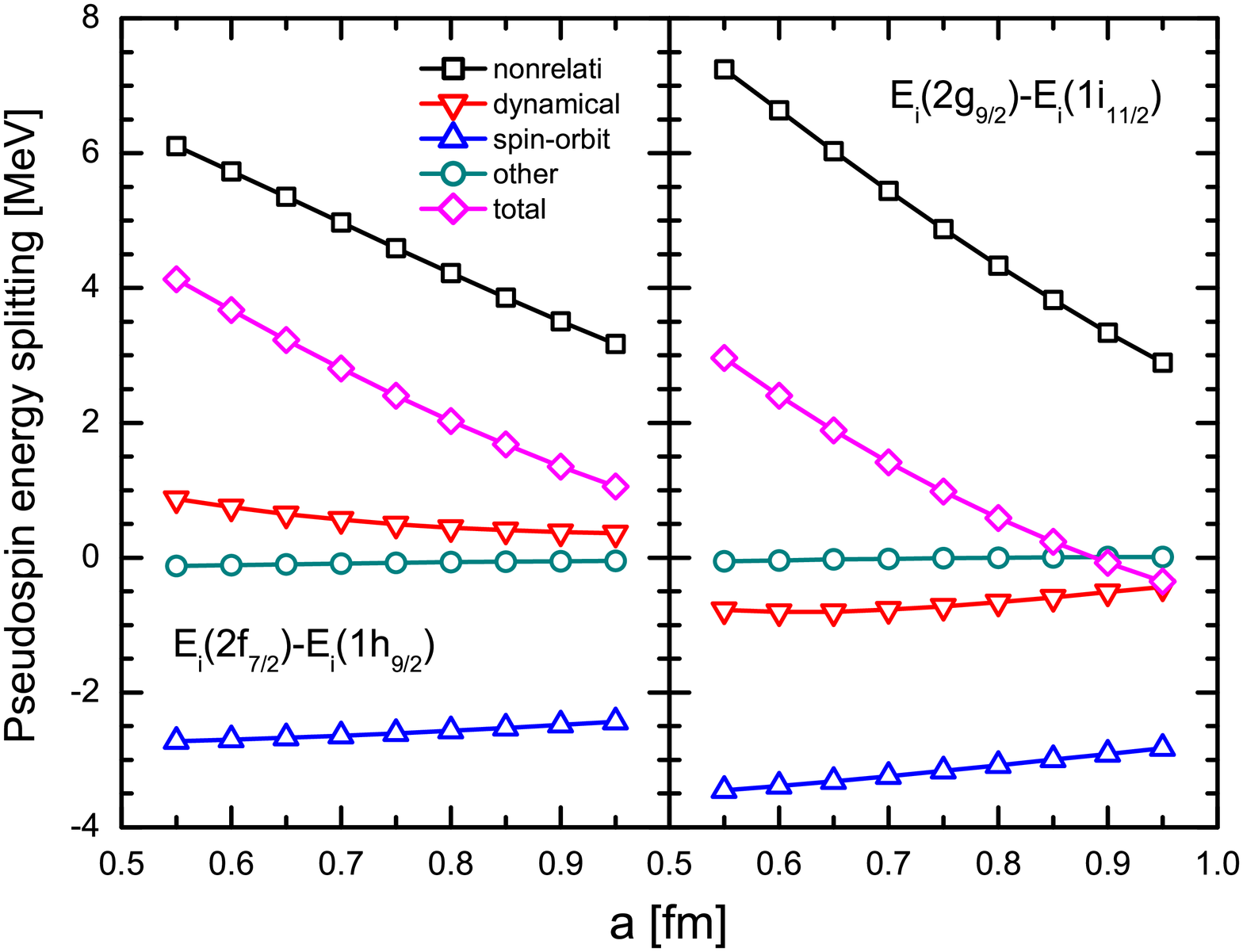}
\caption{(Color online) The same as Fig.4, but for the $(2f_{7/2},1h_{9/2})$
and $(2g_{9/2},1i_{11/2})$ partners.}
\end{figure}

\begin{figure}[tbp]
\includegraphics[width=8cm]{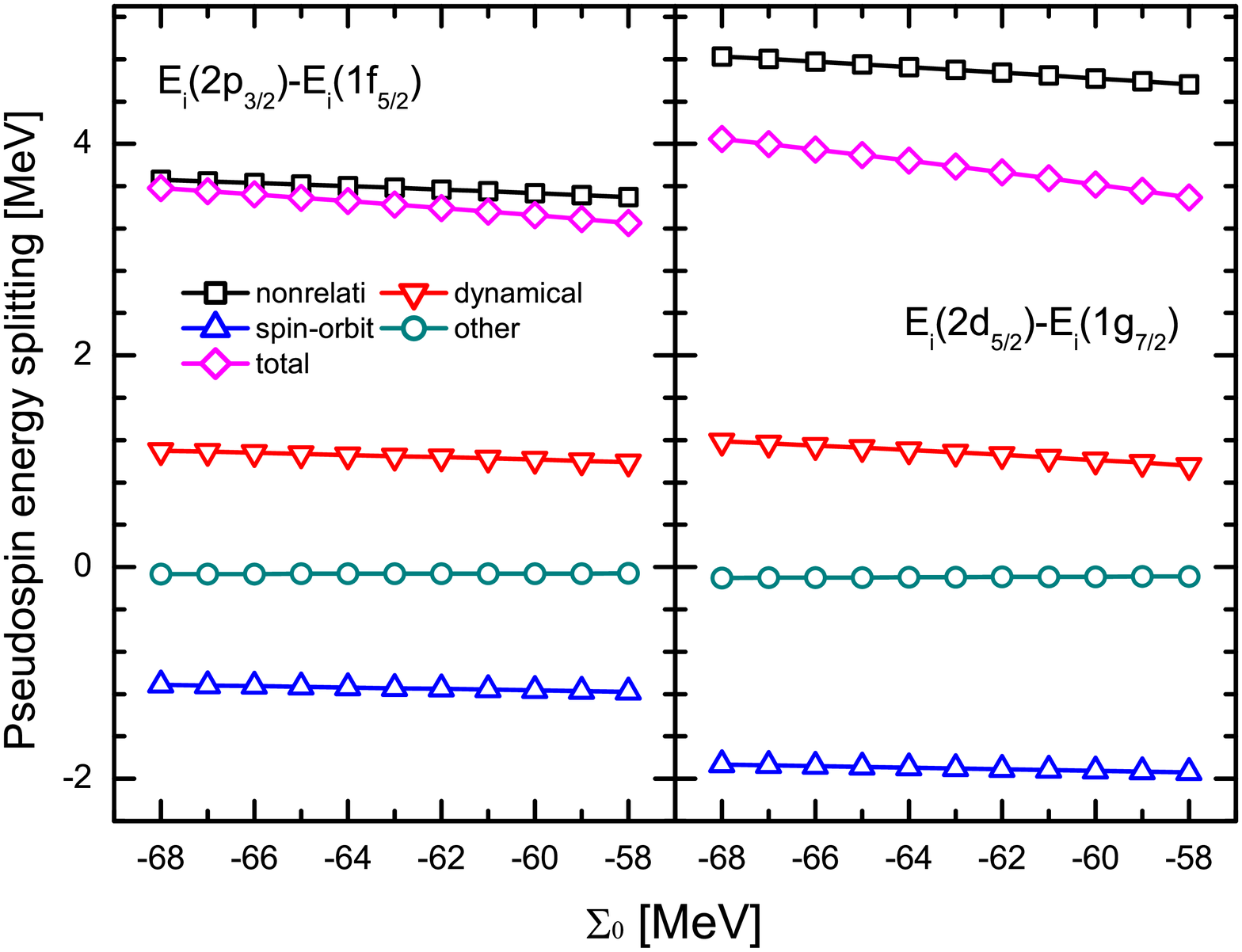}
\caption{(Color online) The same as Fig.4, but with the depth $\Sigma_0$ of
the Woods-Saxon potential for the $(2p_{3/2},1f_{5/2})$ and $%
(2d_{5/2},1g_{7/2})$ partners. }
\end{figure}

\begin{figure}[tbp]
\includegraphics[width=8cm]{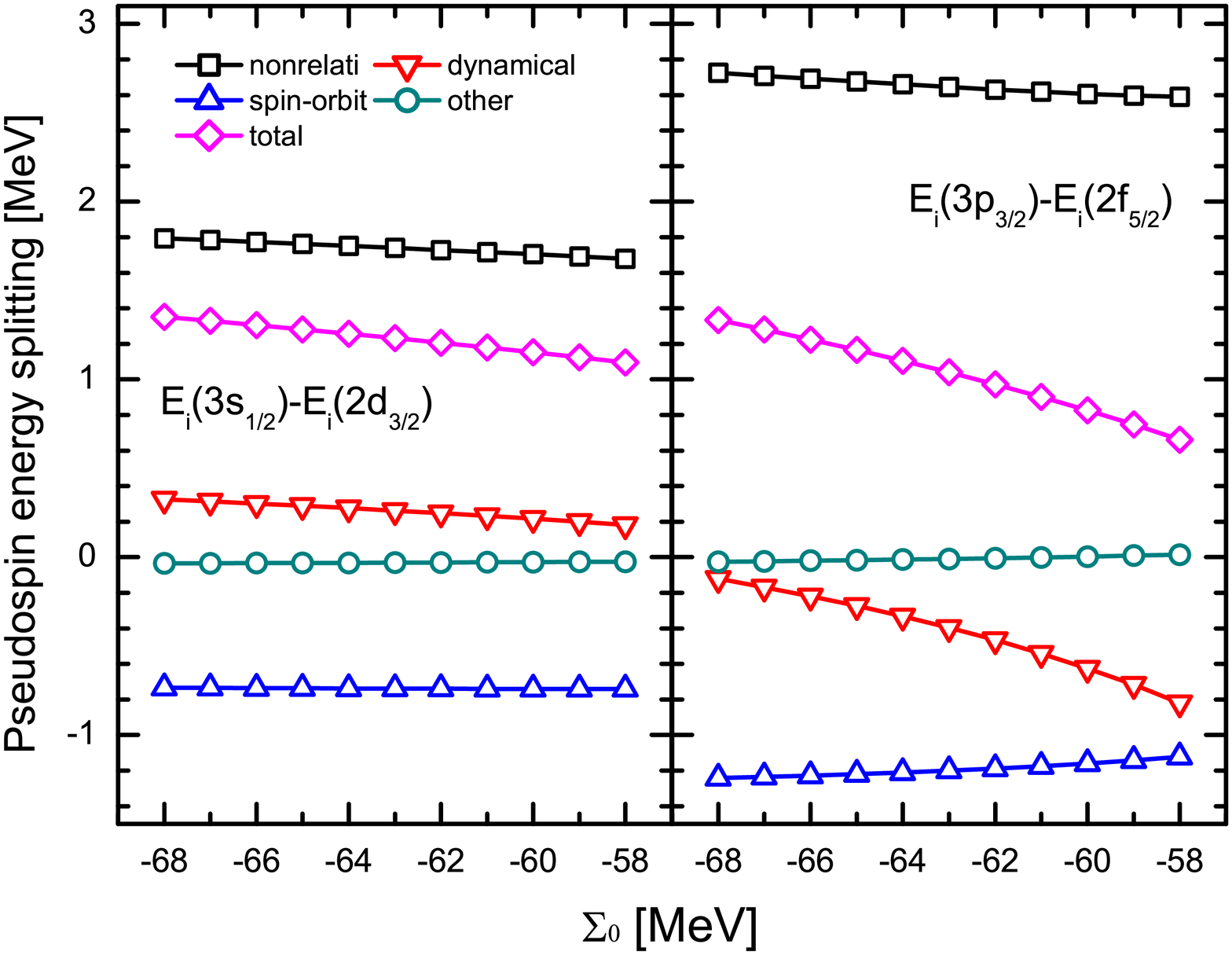}
\caption{(Color online) The same as Fig.7, but for the $(3s_{1/2},2d_{3/2})$
and $(3p_{3/2},2f_{5/2})$ partners. }
\end{figure}

\begin{figure}[tbp]
\includegraphics[width=8cm]{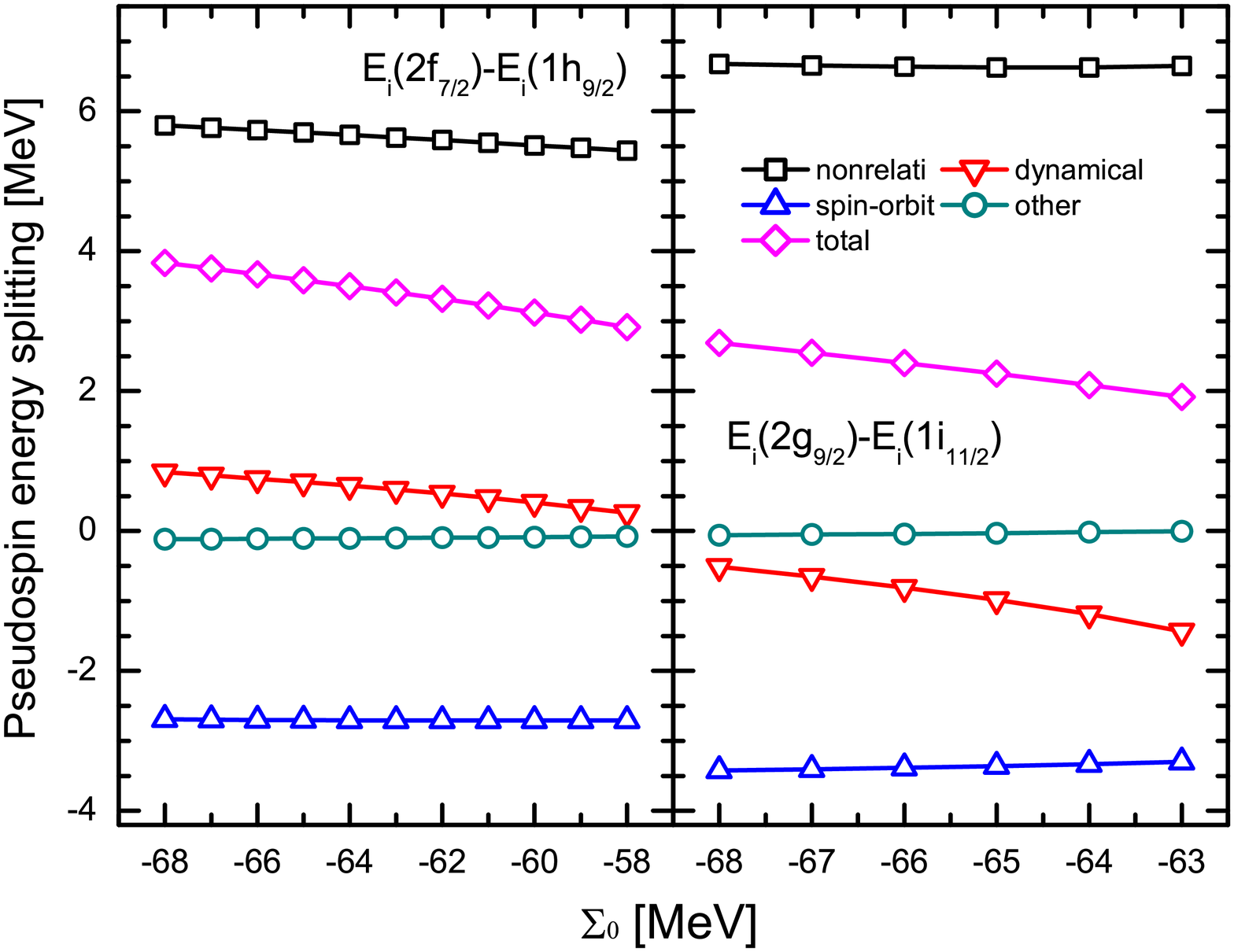}
\caption{(Color online) The same as Fig.7, but for the $(2f_{7/2},1h_{9/2})$
and $(2g_{9/2},1i_{11/2})$ partners. }
\end{figure}

\begin{figure}[tbp]
\includegraphics[width=8cm]{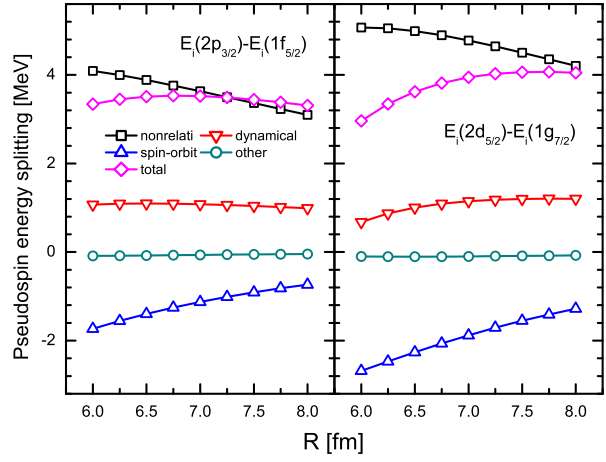}
\caption{(Color online) The same as Fig.4, but with the radius $R$ of the
Woods-Saxon potential for the $(2p_{3/2},1f_{5/2})$ and $(2d_{5/2},1g_{7/2})$
partners.}
\end{figure}

\begin{figure}[tbp]
\includegraphics[width=8cm]{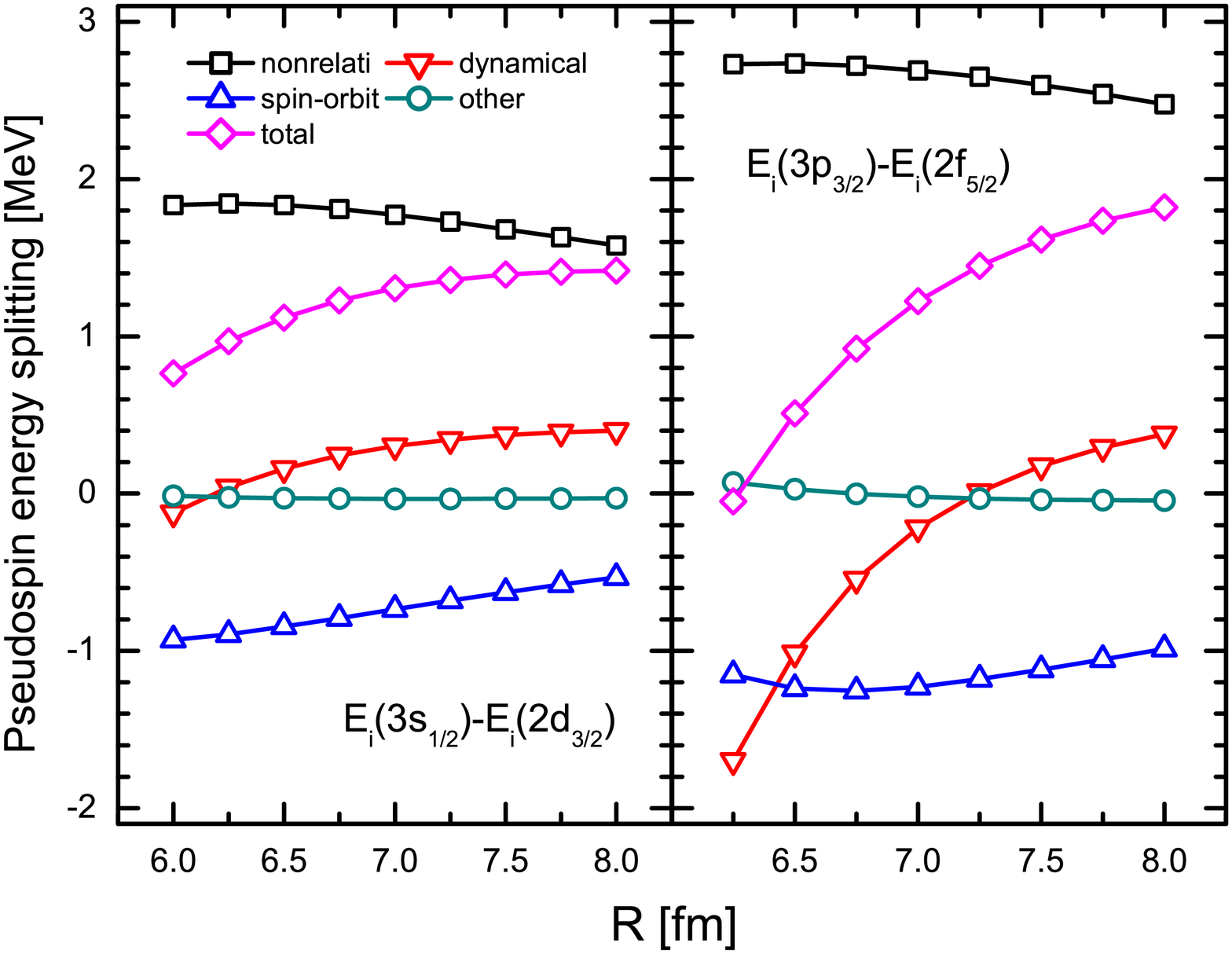}
\caption{(Color online) The same as Fig.10, but for the $(3s_{1/2},2d_{3/2})$
and $(3p_{3/2},2f_{5/2})$ partners. }
\end{figure}

\begin{figure}[tbp]
\includegraphics[width=8cm]{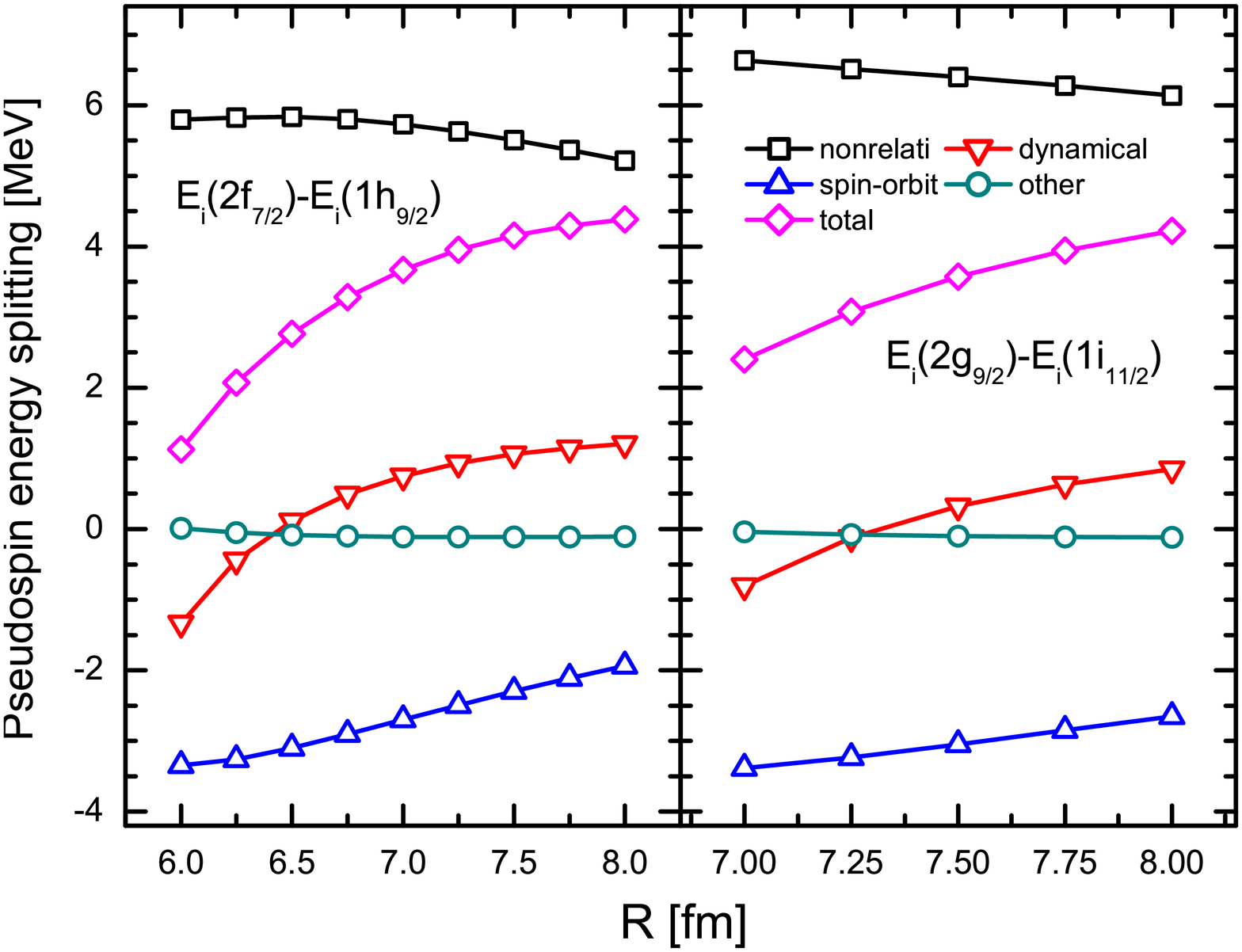}
\caption{(Color online) The same as Fig.10, but for the $(2f_{7/2},1h_{9/2})$
and $(2g_{9/2},1i_{11/2})$ partners. }
\end{figure}

\end{document}